%% file: ms.tex
\documentclass[preprint2,letterpaper]{emulateapj}
\usepackage{natbib}
\bibpunct{(}{)}{;}{a}{}{,}

\begin{document}
\title{Cross-Identification Performance from Simulated Detections: GALEX and SDSS}

\author{S\'ebastien Heinis, Tam\'as Budav\'ari, Alexander S. Szalay}
\affil{Department of Physics and Astronomy, The Johns Hopkins
  University, 3400 North Charles Street, Baltimore, MD 21218}

\begin{abstract}

  We investigate the quality of associations of astronomical sources
  from multi-wavelength observations using simulated detections that
  are realistic in terms of their astrometric accuracy, small-scale
  clustering properties and selection functions. We present a general
  method to build such mock catalogs for studying associations, and
  compare the statistics of cross-identifications based on angular
  separation and Bayesian probability criteria. In particular, we
  focus on the highly relevant problem of cross-correlating the
  ultraviolet GALEX and optical SDSS surveys. Using refined
  simulations of the relevant catalogs, we find that the probability
  thresholds yield lower contamination of false associations, and are
  more efficient than angular separation. Our study presents a set of
  recommended criteria to construct reliable crossmatch catalogs
  between SDSS and GALEX with minimal artifacts.

\end{abstract}


\maketitle

\section{Introduction}

Astrophysical studies can gain significantly by associating data from
different wavelength ranges of the electromagnetic spectrum.
Dedicated multi-wavelength surveys have been a strong focus of
observational astronomy in recent years, e.g. AEGIS
\citep{Davis_2007}, COSMOS \citep{Scoville_2007}, or GOODS
\citep{Dickinson_2003}. At redshifts lower than those probed by these
surveys, several surveys of NASA's Galaxy Evolution Explorer
\citep[GALEX;][]{Martin_2005} essentially provide the perfect
ultraviolet counterparts of the Sloan Digital Sky Survey
\citep[SDSS;][]{York_2000} optical data sets. These surveys or the
combination of these datasets enables to provide invaluable insights
on stars and galaxy properties.



Naturally, these data are taken by different detectors of the separate
projects, hence it is required to combine their information by associating the
independent detections.
Recent work by \citet{Budavari_2008} laid down the statistical
foundation of the cross-identification problem.
Their probabilistic approach assigns an objective Bayesian evidence
and subsequently a posterior probability to each potential
association, and can even consider physical information, such as
priors on the spectral energy distribution or redshift, in addition to
the positions on celestial sphere.
In this paper, we put the Bayesian formalism to work, and aim to
assess the benefit of using posterior probabilities over simple angular
separation cuts using mock catalogs of GALEX and SDSS.
In Section~\ref{sec_simulations}, we present a general procedure to
build mock catalogs that take into account source confusion and
selection functions.
Section~\ref{sec_xmatch} provides the details of
the cross-identification strategy, and defines the relevant
quality measures of the associations based on
angular separation and posterior probability.
In Section~\ref{sec_results}, we present the results for the GALEX-SDSS
cross-identification, and propose a set of criteria to build
reliable combined catalogs.

\section{Simulations}\label{sec_simulations}

The goal is to mimic as close as possible the process of observation
and the creation of source lists.
First, a mock catalog of artificial objects is generated with known
clustering properties, using the method of \citet{Pons-Borderia_1999}.
We then complement this by adding observational effects that are not
included in this method. We generate simulated detections as
observations of the artificial objects with given astronometric
accuracy and selections. Hence the difference between separate sets of
simulated detections, say for GALEX and SDSS, is not only in the
positions, but also they are different subsets of the mock objects.

\subsection{The Mock Catalog}

\begin{figure}[t]
  \plotone{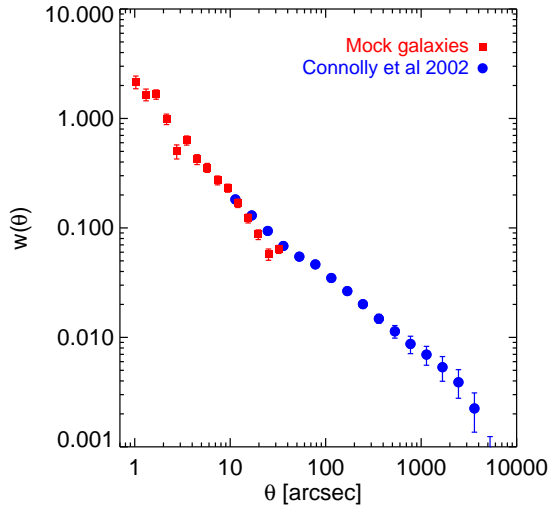}
  \caption{\small Angular correlation function of mock galaxies
  (filled squares) compared to the angular correlation function of
  SDSS galaxies selected with $18<r^{\star}<22$, from
  \citet{Connolly_2002} (filled circles).}
  \label{fig_wtheta}
\end{figure}

We built the mock catalog as a combination of clustered sources (for
galaxies) and sources with a random distribution (for stars). To
simulate clustered sources, we generate a realization of a Cox point
process, following the method described by
\citet{Pons-Borderia_1999}. This point process has a known correlation
function which is similar to that observed for galaxies. We create
such a process within a cone of 1Gpc depth; assuming the notation of
\citet{Pons-Borderia_1999}, we used $\lambda_s = 0.1$ and $l =
1h^{-1}$Mpc for the Cox process parameters. For our purpose, it is
sufficient that the distribution on the sky (i.e., the angular
correlation function) of the mock galaxies displays clustering up to
scales equal to the search radius used for the cross-identification
(5\arcsec~here) and that this distribution is similar to the actual
one. Figure \ref{fig_wtheta} shows the angular correlation function of
our mock galaxy sample (filled squares) along with the measurement
obtained by \citet{Connolly_2002} from SDSS galaxies with
$18<r^{\star}<22$. Note that the galaxy clustering is not well known
at small scales ($\theta < 10\arcsec$) because of the combination of
seeing, point spread function, etc. Hence there is no constraint in
his regime. There is nevertheless a good overall agreement between our
mock catalog and the observations at scales between 10 and 30\arcsec.

In the case of both GALEX and SDSS, galaxies and stars show on average
similar densities over the sky. We create a mock catalog over 100
sqdeg with a total of 10$^7$ sources, half clustered and half
random. The minimum Galactic latitude at which this mock catalog is
representative is around 25$^{\circ}$. For this case study we do not
consider the variation of star density with Galactic latitude; we note
that several mock catalogs can be constructed with different star
densities, and prior probabilities (see sect. \ref{sec_xmatch})
varying accordingly.

\subsection{Simulated Detections}\label{sec_detections}

\begin{figure}[t]
  \plotone{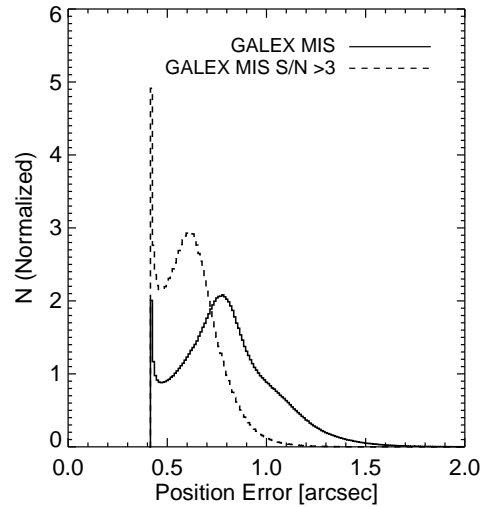}
  \caption{\small Distribution of astrometry errors for simulated
  detections. The solid line shows errors on nuv detections for the
  selection of all GALEX MIS objects, and dotted line for the MIS
  objects with S/N $>3$. These distributions are normalized by their
  integrals.}
  \label{fig_sigmas}
\end{figure}

From our mock catalog we create two sets of simulated detections,
using the approximate astrometry errors of the surveys we consider. We
assume that the errors are Gaussian, and create two detections for
each mock object: a mock SDSS detection with $\sigma_{S}$, and a mock
GALEX one with $\sigma_{G}$. We consider constant errors for SDSS, and
variable errors for GALEX. For GALEX we focus here on the case of the
Medium Imaging Survey (MIS); we will consider two selections: all MIS
objects, or MIS objects with signal-to-noise ratio (S/N) larger than
3. We randomly assign to the mock sources errors from objects of the
GALEX datasets following the relevant selections and using the
position error in the NUV band (\texttt{nuv\_poserr}). The
distributions of these errors are shown on figure \ref{fig_sigmas}. In
the case of GALEX, the position errors are defined as the combination
of the Poisson error and the field error, added in quadrature. The
latter is assumed to be constant over the field (and equal to
0.42\arcsec~in NUV). For SDSS we assume that $\sigma_{S} = 0.1\arcsec$
for all objects. Our results are unchanged if we use variable SDSS
errors for our SDSS mock detections, as the SDSS position errors are
significantly smaller than the GALEX ones.

\subsection{Selection function and confusion}\label{sec_merging}


Our goal is to use the mock catalog described above as a predictive
tool in order to assess the quality of the cross-identifications
between two datasets, here GALEX and SDSS. Hence our mock catalog has
to present similar properties than the data. In practice we need to
include two effects: the selection functions of both catalogs in order
to match the number density of the data, as well as the confusion of
detections caused by the combination of the seeing and point spread
functions.

To apply the selection function, we assign to each mock source a
random number $u$, drawn from an uniform distribution, which
represents the property of the objects. We use the values of $u$ to
select the simulated detections we further consider to study a given
case of cross-identification. The length of the interval in $u$ sets
the density for a given mock catalog. Using the notations of
\citet{Budavari_2008}, we computed the number of SDSS GR7 sources,
$N_{\rm{SDSS}}$ and GALEX GR5, $N_{\rm{}GALEX}$, and scaled them to
the area of our mock catalog. These numbers set the interval in $u$
for both detection sets. We then use the overlap between the intervals
in $u$ to set the density of common objects, as set by the prior
probability determined independently from the data (see sect.\
\ref{sec_xmatch}).

To simulate the confusion of the detections, we performed the
cross-identification of the SDSS and GALEX detections sets with
themselves, using a search radii of 1.5\arcsec~and
5\arcsec~respectively. These values of search radius correspond to the
effective widths of the PSF in both surveys \citep{Stoughton_2002,
  Morrissey_2007}\footnote{see also
  \url{http://www.sdss.org/DR7/products/general/seeing.html}}. We then
consider only the detections that satisfy the selection function
criterion, and merge them. For SDSS, we keep one source chosen
randomly from the various identifications. For GALEX, we keep the
source with the largest position error.

This procedure is repeated for each cross-identification we consider,
as modifying the selection function naturally implies a change in the
number densities and priors.

\section{Cross identification}\label{sec_xmatch}
We performed the cross-identification between the SDSS and GALEX
detection sets using a 5\arcsec~radius. For each association
\citep[see][]{Budavari_2008}, we compute the Bayes factor
\begin{equation}
  B(\psi; \sigma_{S}, \sigma_{G}) = \frac{2}{\sigma^2_{S} +
  \sigma^2_{G}}\exp\left[-\frac{\psi^2}{2(\sigma^2_{S} +
  \sigma^2_{G})}\right]
\end{equation}
where $\psi$ is the angular separation between the two detections, and
is expressed here in radians, as $ \sigma_{S}$ and $\sigma_{G}$. We
also derive the posterior probability that the two detections are from
the same source
\begin{equation}\label{eq_posterior}
  P = \left[1 + \frac{1-P_0}{B\,P_0} \right]^{-1}
  \approx
  \frac{B\,P_0}{1+B\,P_0}
\end{equation}
where $P_0$ is the prior probability, and the approximation is
for the usually small priors.

The Bayes factor, and hence the posterior probability depend on the
position errors from both surveys. As we use a constant prior $P_0$,
this implies that if all objects have the same position errors within
a survey, the posterior probability depends on the angular separation
only. In this case, there is no difference between using a criterion
based upon separation or probability.

We use the posterior probability rather than the Bayes factor as a
criterion. In the assumption of a constant prior probability, the
posterior probability is a monotonic function of the Bayes factor.
However, while we consider here for our case study that the prior is a
constant, in practice it may vary over the sky. Note also that for
instance two surveys with similar position error distributions can
have different priors and then a criterion defined on the basis of the
Bayes factor for one survey can not be applied directly to the other
one.
 
In order to set the overlap between our two detection sets as
described in sect. \ref{sec_merging} to match the selection functions
of the actual datasets, we need to compute the prior $P_0$ from the
data, using the actual cross-identification between GALEX GR5 and SDSS
DR7.

The prior is given by

\begin{equation}\label{eq_prior}
P_0 = \frac{N_{\star}}{N_{SDSS} N_{GALEX}}.
\end{equation}

$N_{\star}$ is the number of sources in the overlap between the
various selections (angular, radial, etc \ldots) of the catalogs
considered for the cross identification, i.e. the number of sources in
the resulting catalog.  We use the self-consistency argument discussed
by \citet{Budavari_2008}
\begin{equation}\label{eq_iter}
\sum P = N_{\star}
\end{equation}
to derive $P_0$. To choose the value of the prior, we use the
iterative process described in \citet{Budavari_2008}. We start the
process by setting $N_{\star} = min(N_{SDSS},N_{GALEX})$.  We then
compute the sum of the posterior probabilities derived using
eq. \ref{eq_posterior}. According to eq. \ref{eq_iter}, this sum gives
us a new value for $N_{\star}$. The same procedure is then repeated
using this updated value, yielding an updated value of the prior as
well.  The chosen value for the prior is obtained after convergence;
we hereafter call this value the \textit{observed} prior.

\begin{figure}[htbp]
  \plotone{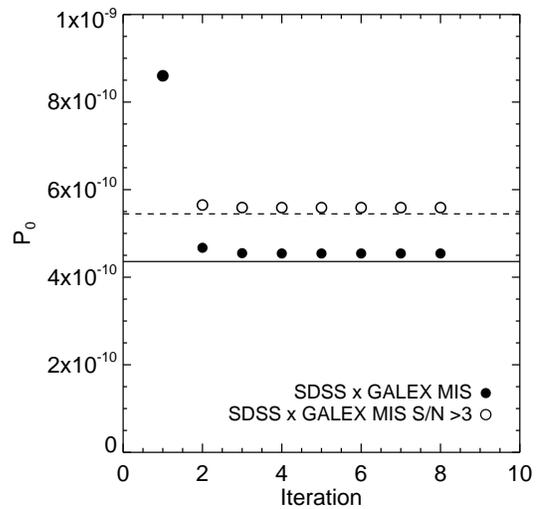}
  \caption{\small Prior probability self-consistent estimation as a
  function of iteration step. Filled circles show the iteration for
  the case of all MIS objects, and open circles for MIS objects with
  S/N $>$3. The solid (dashed) line shows the true prior for all MIS
  objects (MIS objects with S/N $>$3).}
  \label{fig_nstar_iter}
\end{figure}

Then we set the overlap between the two detection sets in our mock
catalog such that the prior value derived for the cross-identification
in the simulations matches the observed one. We use the same iterative
process as described above to determine $P_0$ in the
simulations. Figure \ref{fig_nstar_iter} shows this iteration process
starting from $N_{\star} = N_{\rm{}GALEX}$ for the case with all MIS
objects (filled circles) or MIS S/N $>$3 objects (open circles). The
procedure converges quickly in terms of number of steps. Note also
that the query we use to compute the sum runs in roughly 1 second on
these simulations.

The benefit of the use of simulations is that, in this case, once we
set the overlap between the detection sets required to match the
observed prior, we know the input value of $N_{\star}$ (i.e. the
actual number of detections in the overlap between the two sets) and
hence we can derive the prior corresponding to this number directly
using eq. \ref{eq_prior}, which we call the \textit{true} prior. We
show this true prior on fig. \ref{fig_nstar_iter} as solid line for
the case of all MIS objects, and dashed line for MIS objects with S/N
$>$3. The true priors we are required to use in order to match the
data is slightly lower than the observed ones for both selections: 4\%
lower for all MIS objects and 2.5\% for MIS objects with S/N $>$3. In
other words, we need to use less objects in the overlap between our
detection sets than what we expect from the data.

A different prior value implies a change in the posterior probability;
however the latter also depends on the values of the Bayes factor
$B$. Given the scaling of the relation between the posterior and prior
probabilities (eq. \ref{eq_posterior}), for low $B$ values ($B \ll
1$), a variation of 4\% in the prior yields a variation in posterior
probability of the same amount. For high $B$ values ($B \gg 1$), the
variation is about 0.5\%. Hence this difference between the true and
observed priors has a negligible impact on the values of the posterior
probabilities derived afterwards.

To quantify the quality of the cross-identification, we define the
true positive rate, $T$ and the false positive contamination, $F$. We
can express these quantities as a function of the angular separation
of the association, or the posterior probability. Let $n(x)$ be the
number of associations, where $x$ denote separation or
probability. This number is the sum of the true and false positive
cross-identifications: $n(x) = n_T(x) + n_F(x)$. We define the true
positive rate and false positive contamination as a function of
angular separation as
\begin{eqnarray}
  T(\psi) & = & \frac{\sum n_T(x<\psi)}{N_T}\\
  F(\psi) & = & \frac{\sum n_F(x<\psi)}{\sum n(x<\psi)}
\end{eqnarray}
where $N_T$ is the total number of true associations.
Similar rates are defined as a function of the probability,
\begin{eqnarray}
  T(P) & = & \frac{\sum n_T(x>P)}{N_T}\\
  F(P) & = & \frac{\sum n_F(x>P)}{\sum n(x>P)}.
\end{eqnarray}

\begin{figure}[t]
  \plotone{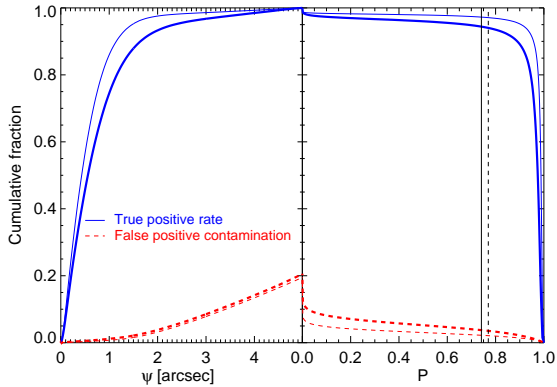}
  \caption{\small True positive contamination rate (in blue, solid
    lines) and false positive contamination (red, dashed lines) as a
    function of angular separation (left) and posterior probability
    (right). GALEX position errors from the full MIS sample yield the
    thick curves; the S/N $>3$ constraint yields the thin curves. We
    also show the posterior probability thresholds defined as in
    \citet{Budavari_2008} (vertical lines on right hand side plot).}
  \label{fig_rates_sep_proba}
\end{figure}

We use the detection merging process to qualify the
cross-identifications as true or false. In our final mock catalog, a
detection represents a set of detections that have been merged. We
therefore consider a case as a true cross-identification where there
is at least one detection in common within the two sets of merged
detections.

Figure~\ref{fig_rates_sep_proba} represents the true positive rate and
the false contamination rate as a function of angular separation
(left) and posterior probability (right). These results suggest that
in the case of the SDSS GALEX-MIS cross-identification, it is required
to use a search radius of 5\arcsec~in order to recover all the true
associations. In the case of all MIS objects, 90\% of the true matches
are recovered at 1.64\arcsec~with a 2.6\% contamination from false
positive. As expected, results are better using objects with high
signal-to-noise ratio (S/N $>$ 3), where 90\% of the true matches are
recovered at 1.15\arcsec~with a 1\% contamination.  Turning to the
posterior probability, the trends are similar to the ones observed as
a function of separation. However, the false positive contamination
increases less rapidly with probability. For instance, a cut at
$P>0.89$ recovers 90\% of the true associations, with a slightly lower
contamination from false positive (2.3\%). We examine in details the
benefits of using separation or probability as a criterion in
Section~\ref{sec_results}.

\section{Results}\label{sec_results}

\subsection{Performance analysis}\label{sec_roc}

\begin{figure}[t]
  \plotone{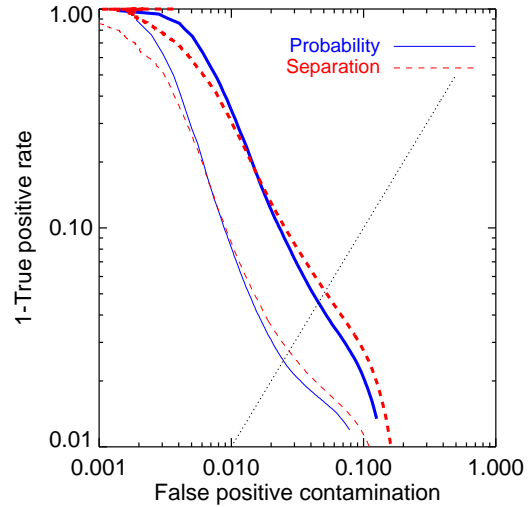}
  \caption{\small Cross identification diagnostic plot: 1-true
    positive rate versus the false positive contamination. These
    quantities are computed as a function of probability (blue, solid
    lines) or separation (red, dashed lines). Thick lines show the
    results for all GALEX MIS objects, and thin lines for GALEX MIS
    objects with S/N $>3$. The dotted line represents the locus of
    $1-T = F$.}
  \label{fig_roc}
\end{figure}

Using the quantities defined above, we can build a diagnostic plot in
order to assess the overall quality of the cross-identification, and
define a criterion to select the objects to use in practice for
further analyses. We show on Fig.\ \ref{fig_roc} the true positive
rate against the false positive contamination, computed as a function
of probability or angular separation. We can compare the false
positive contamination that yields a given true positive rate
threshold for each of these parameters.

The results show that there are some differences between criteria
based on angular separation or posterior probability. Considering all
GALEX MIS objects (solid lines on fig. \ref{fig_roc}), for $1-T
>0.18$, the false contamination rate is slightly lower when using
angular separation as a criterion. This range of true positive rates
corresponds to angular separations smaller than 1.2\arcsec. As there
is a lower limit to the GALEX position errors, this translates into an
upper limit in terms of posterior probability at a given angular
separation. This in turn implies that the probability criterion does
not appear as efficient as separation for associations at small
angular distances in the SDSS-GALEX case.

At $1-T <0.18$, this trend reverses: considering a criterion based on
probability yields a lower false contamination rate.

We can characterize these diagnostic curves by the Bayes threshold,
where $1-T = F$, which minimizes the Bayes error. The location of this
threshold is represented on fig. \ref{fig_roc} by the intersection
between the diagnostic curves and the dotted line. Our results show
that this intersection happens at lower false positive contamination
rate when using the posterior probability as criterion.

For all GALEX objects, the separation where $1-T = F$, $\psi_c$, is
equal to 2.307\arcsec~and the probability, $P_c$ is 0.613. Using the
angular separation as criterion, the Bayes error is then $P_e =
0.102$; using the posterior probability, $P_e = 0.091$. For GALEX
objects with S/N $>$3, $\psi_c = $1.882\arcsec, $P_c$ = 0.665; $P_e =
0.055$ using the angular separation and $P_e = 0.049$ using the
posterior probability.

These results show that a selection based on posterior probability
yields better results (i.e., a lower false contamination rate, and
lower Bayes error) than a selection based on angular separation.

\subsection{Associations}

\begin{figure}[t]
  \plotone{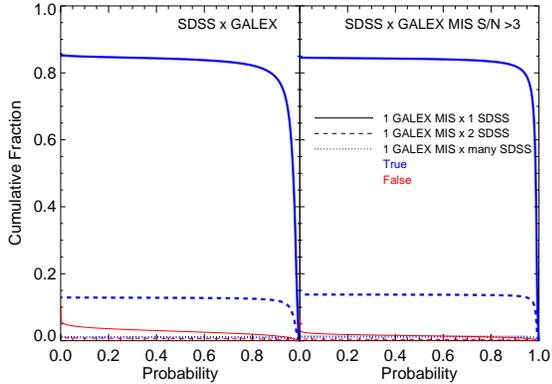}

  \caption{\small True positive rate (blue, thick lines) and false
    contamination rate (red, thin lines) as a function of probability
    for the one GALEX to one SDSS (solid lines), one GALEX to two SDSS
    (dashed lines), one GALEX to many SDSS (dotted lines)
    associations. The left panel show these rates for all GALEX MIS
    objects, and the right one for the GALEX MIS objects with S/N
    $>$3. Note that the curves representing the one GALEX to many SDSS
    associations can barely be seen as the value are too small.}
  \label{fig_rate_proba_1tox}

\end{figure}

\begin{figure}[t]
  \plotone{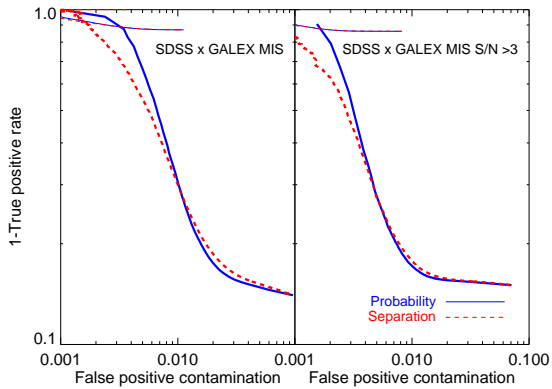}
  \caption{\small 1-True positive rates as a function of the false
    contamination rate for the one GALEX to one SDSS (thick lines) and
    one GALEX to two SDSS (thin lines) associations. The rates are
    computed as a function of probability (blue, solid lines) or
    separation (red, dashed lines). The left panel show these rates
    for all GALEX MIS objects, and the right one for the GALEX MIS
    objects with S/N $>$3.}
  \label{fig_roc_1tox}
\end{figure}

\input{tab1}

Beyond the confused objects, the cross-identification list contains
several types of associations, where a single detection in one catalog
is linked to possibly more than one detection in the other.  We list
in table \ref{tab_xtox} the contingency table of the percentages of
these types in the mock catalog and, in brackets, for the SDSS DR7 to
GALEX GR5 cross-identifications.

The main contribution is from the one GALEX to one SDSS (1G1S, 74\%),
but there are also, for the most significant ones, cases of one GALEX
to two SDSS (1G2S, 21\%) or one GALEX to many SDSS (1GmS,
3\%). Comparing with the data, our mock catalogs are slightly
pessimistic in the sense that the fraction of one to one matches is
lower than in the observations. However, these fractions match
reasonably well enough, which enables us to discuss these cases in the
context of our mock catalogs. We show on figure
\ref{fig_rate_proba_1tox} the true positive and false contamination
rates as a function of probability for the 1G1S (solid lines), 1G2S
(dashed lines), and 1GmS (dotted lines) associations. The 1G1S true
associations represent the bulk (up to 85\%) of the total
cross-identifications. There is also a significant fraction of true
associations within the one 1G2S cases (up to nearly 13\%), while the
1GmS are around 1\%. For the 1G2S or 1GmS cases, we use two methods to
select one object among the various associations: the one
corresponding to the highest probability or the smallest
separation. We computed the true positive and false contamination
rates for these cases as a function of the quantity used for the
selection of the association. We compare the results from these two
methods on figure \ref{fig_roc_1tox}, which shows the diagnostic
curves for the 1G1S (thick lines), 1G2S (thin lines); we do not show
here the 1G2m as they represent only 1\% of the associations. The
diagnostic curves present the same trend than the global ones (see
Fig. \ref{fig_roc}): the posterior probability criterion yields a
lower false contamination rate than the angular separation criterion
above some true positive rate value (e.g., $1-T < 0.29$, for 1G1S
associations considering the cross-identification of all SDSS GALEX
objects).

This is however an artifact caused by the distribution of the GALEX
position errors (see sect. \ref{sec_roc}). For the 1G2S or 1GmS cases,
these results show that true associations can be recovered by
selecting maximal probability, with a low contamination from false
positive (up to around 1\%).

On Figs. \ref{fig_rate_proba_1tox} and \ref{fig_roc_1tox} we compare
the results from all GALEX MIS objects and GALEX MIS objects with S/N
$>$ 3. The quality of the cross-identifications are better for the
latter, for all types of associations.

\subsection{Alternative Error model}

\begin{figure}[t]
  \epsscale{1.}
  \plotone{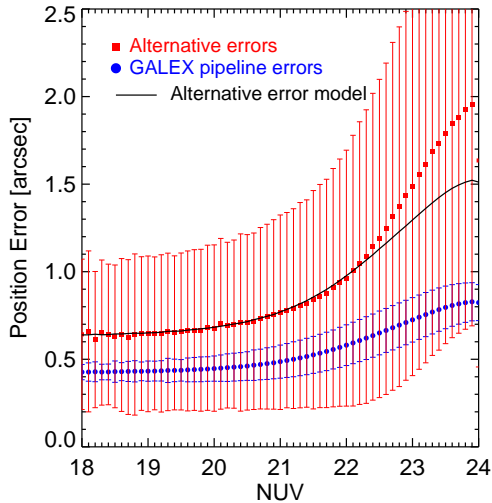}
  \caption{\small GALEX position error as a function of NUV
    magnitude. Circles show the GALEX pipeline error, squares the
    alternative errors (see text). The solid line show the linear
    model we use to modify the GALEX pipeline errors.}
  \label{fig_errors_nuv_mag}
\end{figure}

The accuracy of the analysis of the quality of the
cross-identification strongly depends on the GALEX pipeline position
errors. We use the real data, namely the angular separation to the
SDSS sources measured during the cross-identification between GALEX
GR5 and SDSS DR7, to get an alternative estimation of realistic
errors. In principle the distribution of the angular separations of
the associations depends on the combination of the GALEX and SDSS
position errors. However, the latter are significantly smaller than
the former, so we consider the SDSS errors as negligible here. We
compare on figure \ref{fig_errors_nuv_mag} the dependence on the NUV
magnitude of the position error in the NUV band from the GALEX
pipeline (circles on fig. \ref{fig_errors_nuv_mag}) and the distance
to the SDSS sources (squares), considering only objects classified as
point sources in SDSS. The angular separation between the sources of
the two surveys are significantly larger than the quoted GALEX
pipeline errors. These latter errors are a combination of a constant
field error (equal in NUV to 0.42\arcsec) and a Poisson term. In the
range where both errors estimates are constant ($18<$NUV$<20$), this
comparison suggests that the GALEX field error might be slightly
underestimated. Fot fainter objects, our alternative error increase
faster with magnitude than the GALEX pipeline errors, which might
indicate that this dependence is not well reproduced by the Poisson
term.

We fitted a linear relation to modify the GALEX errors in order to
match the angular separations to the SDSS sources
\begin{equation}
  \textrm{NUV}^{mod}_{poserr} = 2.2\textrm{NUV}_{poserr} - 0.3
\end{equation}
where the position errors are in units of arcsec. This error model is
shown as a solid line on figure \ref{fig_errors_nuv_mag}. It
reproduces well the alternative errors for NUV $\lesssim 22.5$, which
is similar to the 5$\sigma$ limiting magnitude for the MIS in the NUV
band \citep[22.7;][]{Morrissey_2007}.

\begin{figure}[t]
  \epsscale{1.}
  \plotone{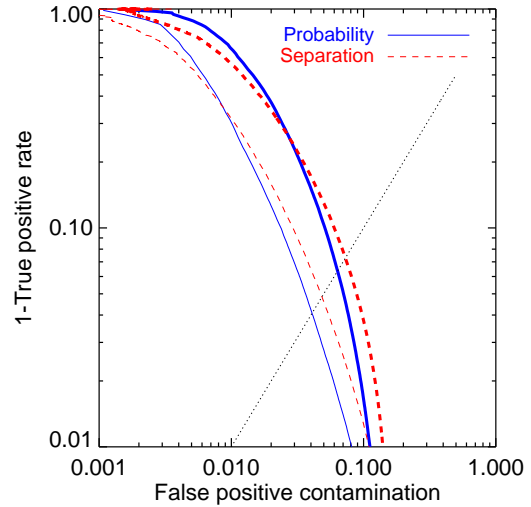}
  \caption{\small Same as figure \ref{fig_roc} using alternative
    position errors for GALEX sources (see text).}
  \label{fig_roc_alt_err}
\end{figure}

We followed the same steps as described in sect. \ref{sec_detections}
and \ref{sec_merging} with these new errors and performed the
cross-identification. The diagnostic curves we obtain are presented on
Fig.~\ref{fig_roc_alt_err}.

The trends are similar to those observed using the GALEX pipeline
errors. The quality of the cross-identification is nevertheless worse
with the alternate errors. In this case, the values of angular
separation and probability where $1-T = F$ are $\psi_c =
3.126\arcsec$, $P_c = 0.711$ for all GALEX objects. Using the angular
separation as a criterion, $P_e = 0.144$ (0.102 with the GALEX
pipeline error), and $P_e = 0.127$ (0.091 with pipeline errors) with
the posterior probability. For GALEX objects with S/N $>$ 3, $\psi_c =
2.514\arcsec$, $P_c = 0.780$ ; $P_e = 0.0958$ (0.055, pipeline errors)
using angular separation, and $P_e = 0.0812$ (0.049, pipeline errors)
with the posterior probability.

In other words, the contamination from false positive is larger at a
given true positive rate. For instance, for all GALEX MIS objects,
with 90\% of the true associations and considering posterior
probability as a criterion, the contamination is 5\% compared to 2.3\%
using the GALEX pipeline errors. Note also that the difference between
the angular separation and the probability diagnostic curves is larger
with this alternate error model. This suggests that the probability is
a more efficient way than angular separation to select
cross-identifications for surveys with larger position errors.

\subsection{Building a GALEX-SDSS catalog}
\input{tab2}

The combination of the results we presented can be used to define a
set of criteria for constructing a reliable joint GALEX-SDSS
catalog. It is natural to have different selections for each type of
association. We will here focus on the 1G1S and 1G2S cases, as they
represent around 95\% of the associations.

In Table \ref{tab_crit} we propose a set of criteria, based on the
posterior probability, to get 90\% of the true cross-identifications,
consisting of 80\% of 1G1S and 10\% of 1G2S. We also list the
corresponding false positive contamination. These cuts enable to build
catalogs with 1.8\% of false positives when using all GALEX objects,
or 0.8\% when using GALEX objects with S/N $>3$.

\section{Conclusions}
We presented a general method using simple mock catalogs to assess the
quality of the cross-identification between two surveys which takes
into account the angular distribution and confusion of sources, and
the respective selection functions of the surveys. We applied this
method to the cross-identification of the SDSS and GALEX sources. We
used the probabilistic formalism of \citet{Budavari_2008} to study how
the quality of the associations can be quantified by the posterior
probability. Our results show that criteria based on posterior
probability yield lower contamination rates from false positive than
criteria based on angular separation. In particular, the posterior
probability is more efficient than angular separation for surveys with
larger position errors. Our study also suggest that the GALEX pipeline
position errors might be underestimated and we described an
alternative measure of these errors. We finally proposed a set of
selection criteria based on posterior probability to build reliable
SDSS-GALEX catalogs that yield 90\% of the true associations with less
than 2\% contamination from false positives.

\section{Acknowledgements}
The authors gratefully acknowledge support from the following
organisations: Gordon and Betty Moore Foundation (GMBF 554),
W. M. Keck Foundation (Keck D322197), NSF NVO (AST 01-22449), NASA
AISRP (NNG 05-GB01G), and NASA GALEX (44G1071483).

\end{document}

%% file: tab1.tex
\begin{deluxetable}{cccc}
\tablecolumns{4} \tabletypesize{\footnotesize} \tablewidth{0pt}
\tablecaption{\small Percentages of associations by type \label{tab_xtox}}
\tablehead{
\multicolumn{4}{c}{SDSS}\\
\colhead{GALEX} & \colhead{1} &\colhead{2} & \colhead{Many}
} 
\startdata 
1 & 74.061 (75.870) & 21.007 (18.595) & 2.577 (2.469)\\[0.1cm]
2 & 1.146   (2.253) & 1.006 (0.697)  & 0.188 (0.102)\\[0.1cm]
Many & 0.006 (0.009) & 0.007 (0.004) & 0.002 (0.001)\\[0.1cm]
\enddata

\tablecomments{ Percentages of associations by type in the mock
catalogs. The numbers in brackets give the percentages from the
cross-identification of SDSS DR7 and GALEX GR5 data. All percentages
are given with respect to the total number of matches.}
\end{deluxetable}

%% file: tab2.tex
\begin{deluxetable*}{ccc}
\tablecolumns{3} \tabletypesize{\footnotesize} \tablewidth{0pt}
\tablecaption{\small Selection criteria for SDSS-GALEX sample\label{tab_crit}}
\tablehead{\colhead{Association} & \colhead{Probability cut} & \colhead{False positive contamination}
} 
\startdata 
1 GALEX to 1 SDSS & $P>0.877$  & 1.6\\[0.1cm]
1 GALEX to 2 SDSS & $P>0.955$   & 0.2\\[0.1cm]
1 GALEX (S/N $>$ 3) to 1 SDSS & $P>0.939$  & 0.7\\[0.1cm]
1 GALEX (S/N $>$ 3) to 2 SDSS & $P>0.982$   & 0.1\\[0.1cm]
\enddata

\tablecomments{Posterior probability cuts to obtain 80\% (10\%) of the
true associations for the one GALEX to one SDSS (one GALEX to two
SDSS) matches. The corresponding false positive contamination
percentages are also listed. The first two lines give the cuts for all
GALEX MIS objects and the two last ones for the GALEX MIS objects with
S/N $>$ 3.}
\end{deluxetable*}